12 July 2021

# The Gaia Successor in 2021

*Erik Høg, Niels Bohr Institute, Copenhagen University, Denmark.* ehoeg@hotmail.dk.

**Abstract:** A new mission about twenty years after Gaia with similar astrometric performance would be important for all branches of astronomy. The two missions together would, e.g., give much more accurate motions of the common objects due to the large epoch difference. By adding a Near-InfraRed (NIR) capability to the new mission we will be able to peer into the obscured regions of the Galaxy and measure up to 10 or 12 billion new objects and reveal many new sciences in the process. ESA has now ranked the development of this mission so high that a launch about 2045 is quite probable. A brief history of the project is included.

## The Project Begin

ESA had called for new Large missions. I submitted my response[1] *"Astrometry for Dynamics"* (Høg 2014a) in May 2013, before the launch of Gaia, and then worked on the project, alone for the first two years. It was clear to me from the beginning that a mission twenty years beyond Gaia would be important for all branches of astronomy: For studies of stars, interstellar matter, the Milky Way, other galaxies, quasars, dark matter and of our own planetary system and exo-planets. But it was also clear that much work was needed to bring that message out.

Thanks to the response from a hundred competent people in oral discussion or correspondence, a draft report from 2013 was updated time after time during four years (Høg 2014b). Many supplementing reports were issued as listed in the references of the report, and I held numerous lectures on this matter.

## Status in June 2021

First to be noted, Gaia has in July 2021 been observing for seven years and is expected to continue for three or more years. The third data release EDR3 (ESA 2020) on 3 Dec. in 2020 with data for 1.8 billion sources was based on 34 months of observation.

Very exciting these days is the **Gaia successor in twenty years**, see the message from David Hobbs in the appendix. ESA has decided to continue the development of such a satellite considering its scientific importance. The decision is at such a level that a launch in about twenty years is quite probable. The satellite shall have sensitivity in the visual but also in the near infrared and is therefore called GaiaNIR. It shall obtain astrometry and photometry in the visual and NIR in order to reach very cold red stars and stars reddened by interstellar clouds. Altogether, about 10-12 billion stars and quasars are expected including the 2 billion being observed by Gaia.

---

[1] Some comments to my response in Høg (2014a) are worth mentioning. In the acknowledgements, Lennart Lindegren is quoted: "I strongly encourage you to submit the proposal, which I think is definitely a very viable concept" and "The argument for high angular resolution photometry is compelling, for it is certainly one of the weak points of Gaia." Two other comments from prominent astronomers were far from positive, one of them wrote that such a proposal as mine should only be submitted by a committee, not by a single person, but I had no time to do so as I explain in the pre-ample. The other comment was that my proposal was not ripe and should never have been submitted, and that the same was true for the other proposal which was for sub-microarcsecond astrometry submitted by a very competent team. After all, ESA decided not to pursue any of the two proposals at that time, but my proposal put me on the track to continue the work.

The project started with my proposal to ESA in May 2013 and in 2014 the quest for an infrared capability began. This was intensified in 2015 when David Hobbs in Lund became interested, and in 2017 we obtained an ESA study of the mission. Especially the NIR detector is critical since we need TDI, time-delayed integration, which was easy to do in the visual with CCDs in Gaia but is difficult in NIR. An industrial study of NIR detection is ongoing. In 2019 we began a quest for significant international contribution to a mission with ESA leading. Together with US, Japanese and Australian colleagues two proposals for study have been submitted. International contribution is also quite natural because of the great importance for astronomy world wide of this powerful all-sky high-accuracy astrometric mission, see Høg & Hobbs 2019.

The project is in good hands with David Hobbs, not depending any longer on an old guy who became 89 recently, but I am still writing and giving invited talks online. David ends his message saying: *"It is clear we must now embark on a long proposal process and make ourselves very visible to sell our ideas." - GOOD LUCK ! - and great thanks to David.*

**References**
Høg E. 2014a, **Astrometry for Dynamics**. Submitted to ESA in May 2013 as White paper proposal for a Large mission. http://arxiv.org/abs/1408.3299

Høg E. 2014b, **Absolute astrometry in the next 50 years**. Report with updates until 12 June 2017.  36 pages http://arxiv.org/abs/1408.2190 , https://ui.adsabs.harvard.edu/#abs/arXiv:1408.2190

Høg, E., Hobbs, D.  2019,  **Gaia Successor with International Participation.** Proceedings of the symposium *Journées 2019,* Astrometry, Earth Rotation, and Reference Systems in the GAIA era, in Paris on Oct. 7-9, p. 49-53, edited by C. Bizouard. At http://www.astro.ku.dk/~erik/xx/ParisHoeg2019.pdf

ESA 2020, **Gaia Early Data Release 3 (Gaia EDR3).** https://www.cosmos.esa.int/web/gaia/earlydr3 .

# Appendix
## GaiaNIR and ESA in 2021

**Erik Høg**  in a mail on 12-06-2021

Very good news for GaiaNIR has come from the ESA Senior Committee according to the following mail from David Hobbs today.
-----------------------------------
**David Hobbs**

To 24 recipients who co-authored the proposal to ESA:

Dear all,

The full report is available here
https://www.cosmos.esa.int/documents/1866264/1866292/Voyage2050-Senior-Committee-report-public.pdf/e2b2631e-5348-5d2d-60c1-437225981b6b?t=1623427287109
and it clarifies a lot.

**For L-class missions** we are in competition with LIFE which is given higher scientific priority.
https://www.life-space-mission.com

However LIFE requires more challenging technological developments as a formation flying nulling interferometer. The report recommends that studies be conducted to see if it can fit in an L-class mission budget. I presume this is similar to the CDF study we have already undergone.

We are mentioned explicitly in the report as "The Galactic Ecosystem with Astrometry in the Near-infrared". I see our inclusion as an L-class concept as a major success - our proposal was effectively as an M-class mission with international collaboration to make it affordable.

**For M-class missions** we are also mentioned under "High Precision Astrometry" and here we seem to have a higher priority than pointed relative astrometry.

"Global astrometry in the near IR as described in Section 2.2.2 would have a much broader impact…"

This reflects our original proposal as an M-class mission with international collaboration.

For the immediate future we have to follow up the detector developments as they are critical to achieving the science goals. US NIR detectors are still mentioned as an example which means the rotating mirror concept is still being pushed by ESA even though it gives a weaker scientific return. I still feel the NIR-APD detectors show greater promise but let us see.

In summary we have a shot at both M- and L-class missions. It is clear we must now embark on a long proposal process and make ourselves very visible to sell our ideas.
In the end I hope that both LIFE and our NIR mission will fly as both have compelling scientific goals.

All the best
David

David Hobbs
Associate Professor (Universitetslektor)
Lund Observatory, Sweden